\input{aipcheck.tex}

  \def\selectedoptions{final}

\documentclass[
   \selectedoptions
  ]
  {aipproc}

\layoutstyle{6x9}

\SetInternalRegister\hbadness{8000} 

\newcommand\doingARLO[2][]{%
  \ifx\mmref\undefined #1\else #2\fi
}
\begin{document}
\newcommand{\bi}{\begin{itemize}}
\newcommand{\ei}{\end{itemize}}
\newcommand{\be}{\begin{equation}}
\newcommand{\ee}{\end{equation}}
\newcommand{\ba}{\begin{eqnarray}}
\newcommand{\ea}{\end{eqnarray}}
\newcommand{\bse}{\begin{subequations}}
\newcommand{\ese}{\end{subequations}}
\newcommand{\M}{{\cal {M}}}
\newcommand{\C}{{\cal {C}}}
\newcommand{\CS}{{\cal {S}}}
\newcommand{\la}{\langle}
\newcommand{\ra}{\rangle}
\newcommand{\kB}{k_{_B}}
\newcommand{\vT}{v_{_T}}
\newcommand{\vlim}{v_{\hbox{\tiny{lim}}}}
\newcommand{\rhoha}{\rho^{\hbox{\tiny{(h)}}}}
\newcommand{\rhohac}{\rho_c^{\hbox{\tiny{(h)}}}}
\newcommand{\nha}{n^{\hbox{\tiny{(h)}}}}
\newcommand{\pha}{p^{\hbox{\tiny{(h)}}}}
\newcommand{\sha}{s^{\hbox{\tiny{(h)}}}}
\newcommand{\shac}{s_c^{\hbox{\tiny{(h)}}}}
\newcommand{\shacth}{s_c^{\hbox{\tiny{(h)}}}|_{_{\hbox{\tiny{th}}}}}
\newcommand{\shacmb}{s_c^{\hbox{\tiny{(h)}}}|_{_{\hbox{\tiny{MB}}}}}
\newcommand{\shacem}{s_c^{\hbox{\tiny{(h)}}}|_{_{\hbox{\tiny{em}}}}}
\newcommand{\nhac}{n_c^{\hbox{\tiny{(h)}}}}
\newcommand{\xhac}{x_c^{\hbox{\tiny{(h)}}}}
\newcommand{\Tha}{T^{\hbox{\tiny{(h)}}}}
\newcommand{\bha}{\beta^{\hbox{\tiny{(h)}}}}
\newcommand{\xha}{x^{\hbox{\tiny{(h)}}}}
\newcommand{\sigha}{\sigma_{\hbox{\tiny{(h)}}}}
\newcommand{\seqq}{s^{\hbox{\tiny{(eq)}}}}
\newcommand{\neqq}{n^{\hbox{\tiny{(eq)}}}}
\newcommand{\Yeqq}{Y^{\hbox{\tiny{(eq)}}}}
\newcommand{\svir}{s_{\hbox{\tiny{(vir)}}}}
\newcommand{\zvir}{z_{\hbox{\tiny{(vir)}}}}
\newcommand{\xvir}{x_{\hbox{\tiny{(vir)}}}}
\newcommand{\xf}{x_{\hbox{f}}}
\newcommand{\nf}{n_{\hbox{f}}}
\newcommand{\sff}{s_{\hbox{f}}}
\def\chic#1{{\scriptscriptstyle #1}}
\title 
      []
      {A New Method to Infere which Type of Neutralinos make up 
Galactic Halos}
%
\author{Luis G. Cabral-Rosetti$^a$, Xabier Hern\'andez$^b$ and 
Roberto A. Sussman$^{a}$}
{address={$^a${\it Instituto de Ciencias Nucleares, Universidad Nacional 
Autónoma de México (ICN-UNAM),\\
Circuito Exterior, C.U., Apartado Postal 70-543, 04510 México, D.F.,  
México.}\\
{$^b${\it Instituto de Astronom{\'\i}a, Universidad Nacional Autónoma 
de México, (IA-UNAM).\\
Circuito de la Investigación Científica, C.U., Apartado Postal 
70-264, 04510 México, D.F.,  México.}
}\\
},
  email={luis@nuclecu.unam.mx, matias@fenix.ifisicacu.unam.mx, 
rosado@fenix.ifisicacu.unam.mx},
  thanks={}
}

\copyrightyear  {2001}

\begin{abstract}
Applying the microcanonical definition of entropy to a weakly interacting 
and self--gravitating neutralino gas, we evaluate the change in the local 
entropy per particle of this gas between the freeze out era and present 
day virialized halo structures. An ``entropy consistency'' criterion emerges 
by comparing the obtained theoretical entropy per particle of the virialized 
halos with an empirical entropy per particle given in terms of dynamical 
halo variables of actual galactic structures.  We apply this criterion to the 
cases when neutralinos are mostly B-inos and mostly Higgsinos, in conjunction 
with the usual ``abundance'' criterion requiring that present neutralino 
relic density complies with $0.2<\Omega_{{\tilde\chi^1_0}} < 0.4$ for 
$h\simeq 0.65$.  The joint application of both criteria reveals that a much 
better fitting occurs for the B-ino than for the Higgsino channels, so that 
the former seems to be a favored channel along the mass range  of
$150\,\hbox{GeV} <  m_{{\tilde\chi^1_0}} < 250 \,\hbox{GeV}$. These results 
are consistent with neutralino annihilation patterns that emerge from recent 
theoretical analysis on cosmic ray positron excess data reported by the HEAT 
collaboration. The suggested methodology can be applied to test other 
annihilation channels of the neutralino, as well as other particle candidates 
of thermal WIMP gas relics \footnote{Contribution to Proceeedings of the 
{\it Fifth Mexican School (DGFM): The Early Universe and Observational 
Cosmology} de la Divisi\'on de Gravitaci\'on y F{\'\i}sica Matem\'atica de 
la Sociedad Mexicana de F{\'\i}sica (DGyFM-SMF). November 24-29 2002
Playa del Carmen, Quitana-Roo, M\'exico. This paper is based on the Ref. 
\cite{nosotros}.} .   
\end{abstract}
%
%
\maketitle

\section{Introduction.}
\label{sec:1}

There are strong theoretical arguments favoring lightest supersymmetric 
particles (LSP) as making up the relic gas that forms the halos of actual 
galactic structures. Assuming that {\it R} parity is conserved and that the 
LSP is stable, it might be an ideal candidate for cold dark matter (CDM), 
provided it is neutral and has no strong interactions. The  most favored 
scenario \cite{Ellis,Report,Torrente,Roszkowski,Fornengo,Ellis2} considers 
the LSP to be the lightest neutralino ($\tilde\chi_1^0$), a mixture of 
supersymmetric partners of the photon, $Z$ boson and neutral Higgs boson 
\cite{Report}. Since neutralinos must have decoupled once they were 
non-relativistic, it is reasonable to assume that they constituted 
originally a Maxwell-Boltzmann (MB) gas in thermal equilibrium with other 
components of the primordial cosmic plasma. In the present cosmic era, 
such a gas is practically collision--less and is either virialized in 
galactic and galactic cluster halos, in the process of virialization or 
still in the linear regime for superclusters and structures near the scale 
of homogeneity \cite{KoTu,Padma1,Peac}.

Besides the constraint due to their present abundance as main constituents 
of cosmic dark matter ($\Omega_{{\tilde\chi^1_0}} \sim 0.3$), it is still 
uncertain which type of annihilation cross section characterizes these 
neutralinos. In this paper we present a method that discriminates between 
different cross sections, based on demanding (together with the correct 
abundance) that a theoretically estimated entropy per particle matches an 
empiric estimate of the same entropy, but constructed with dynamic variables 
of actual halo structures. The application of this ``entropy consistency'' 
criterion is straightforward because entropy is a state variable that can 
be evaluated at equilibrium states, irrespectively of how enormously 
complicated could be the evolution between each state. In this context, the 
two fiducial equilibrium states of the neutralino gas are (to a good 
approximation) the decoupling (or ``freeze out'') and their present state 
as a virialized relic gas. Considering simplified forms of annihilation 
cross sections. the joint application of the abundance and 
entropy--consistency criteria favors the neutralinos as mainly ``B--inos'' 
over neutralinos as mainly ``higgsinos''. These results are consistent with 
the theoretical analysis of the HEAT experiment~\cite{HEAT-TH,HEAT1,HEAT2} 
which aims at relating the observed positron excess in cosmic rays with a 
possible weak interaction between neutralinos and nucleons in galactic halos.

The paper is organized as follows. In section 2 we describe the thermodynamics 
of the neutralino gas as it decouples. Section 3 applies to the 
post--decoupling neutralino gas the entropy definition of the microcanonical 
ensemble entropy, leading to a suitable theoretical estimate of the entropy 
per particle. In section 4 we obtain an empiric estimate of this entropy 
based on actual halo variables, while in section 5 we examine the consequences 
of demanding that these two entropies coincide. Section 6 provides a summary 
of these results.    

\section{The neutralino gas}

The equation of state of a non-relativistic MB neutralino gas is
\cite{KoTu,Padma1,Peac}
%
\ba 
\rho \ =&& \
m_{{\tilde\chi_0^1}}\,n_{{\tilde\chi_0^1}}\,
\left(1+\frac{3}{2\,x}\right),\qquad p \ = \
\frac{m_{{\tilde\chi_0^1}}\,n_{{\tilde\chi_0^1}}}{x}, 
\label{eqst}\\
 x \ \equiv&& \
\frac{m_{{\tilde\chi_0^1}}}{T},
\label{beta_def}
\ea
%
where $m_{{\tilde\chi_0^1}}$ and $n_{{\tilde\chi_0^1}}$ are the neutralino
mass and number density. Since we will deal exclusively with the lightest 
neutralino, we will omit henceforth the subscript $_{{\tilde\chi_0^1}}$, 
understanding that all usage of the term ``neutralino'' and all symbols of 
physical and observational variables ({\it{i.e.}} $\Omega_0,\,m,\,\rho,\,n,$ 
etc.) will correspond to this specific particle. As long as the neutralino 
gas is in thermal equilibrium, we have
\ba 
n \ \approx \ \neqq \ =&& \
g\,\left[\frac{m}{\sqrt{2\,\pi}}\right]^3\,x^{-3/2}\,\exp\,
\left(-x\right),
\label{n_theq}
\ea        
where $g=1$ is the degeneracy factor of the neutralino species. The number 
density $n$ satisfies the Boltzmann equation \cite{Report,KoTu}
\ba 
\dot n + 3\,H\,n \ = \ -\la
\sigma|\hbox{v}|\ra\left[n^2-\left(\neqq\right)^2\right],
\label{boltz}
\ea
where $H$ is the Hubble expansion factor and $\la \sigma|\hbox{v}|\ra$ is 
the annihilation cross section.  Since the neutralino is non-relativistic 
as annihilation reactions ``freeze out'' and it decouples from the radiation 
dominated cosmic plasma, we can assume for $H$ and $\la\sigma|\hbox{v}|\ra$ 
the following forms
\ba 
H \ = \ 1.66\, g_*^{1/2}\frac{T^2}{m_p},
\label{eqH}\\
\la\sigma|\hbox{v}|\ra \ = \ a \ + \ b\la \hbox{v}^2\ra,
\label{eq<sv>}
\ea 
where $m_p=1.22\times 10^{19}$ GeV is Planck's mass, $g_*=g_*(T)$ is the sum 
of relativistic degrees of freedom, $\la \hbox{v}^2\ra$ is the thermal 
averaging of the center of mass velocity (roughly $\hbox{v}^2\propto 1/x$ in 
non-relativistic conditions) and the constants $a$ and $b$ are determined by 
the parameters characterizing specific annihilation processes of the
neutralino (s-wave or p-wave) \cite{Report}. The decoupling of the neutralino 
gas follows from the condition
\ba 
\Gamma \ \equiv \ n\,\la\sigma|\hbox{v}|\ra \ = \ H,
\label{fcond}
\ea
leading to the freeze out temperature $T_{\hbox{f}}$. Reasonable approximated 
solutions of (\ref{fcond}) follow by solving for $x_f$ the implicit relation 
\cite{Report}
\ba 
\xf  = \ln\left[\frac{0.0764\,m_p\,c_0(2+c_0)\,(a+6\,b/\xf)\,m}
{(g_{*{\hbox{f}}} \,\xf)^{1/2}}\right],
\label{eqxf}
\ea  
where $g_{*{\hbox{f}}}=g_*(T_{\hbox{f}}) $ and $c_0\approx 1/2$ yields the 
best fit to the numerical solution of (\ref{boltz}) and (\ref{fcond}). From 
the asymptotic solution of (\ref{boltz}) we obtain the present abundance of 
the relic neutralino gas \cite{Report}
%
\ba 
\Omega_0\,h^2 \ = &&\ Y_\infty\, \frac{\CS_0\, m} 
{\rho_{\hbox{crit}}/h^2}
\ \approx \ 2.82\times 10^8\,Y_\infty\,\frac{m}{\hbox{GeV}},
\label{eqOmega0}\\
 Y_\infty \
\equiv  &&\ \frac{n_0}{\CS_0} \nonumber\\ = && 
\left[0.264\,g_{*\hbox{f}}^{1/2}\,m_p\,m\left\{a/\xf+3(b-1/4\,a)
/\xf^2\right\}\right]^{-1},
\nonumber
\\
\label{eqYinf}
\ea
%
where $\CS_0\approx 4000\,\hbox{cm}^{-3}$ is the present radiation entropy
density (CMB plus neutrinos),  $\rho_{\hbox{crit}} = 1.05 \times
10^{-5}\,\hbox{GeV}\,\hbox{cm}^{-3}$. 

Since neutralino masses are expected to be in the range of tens to hundreds 
of GeV's and typically we have $\xf\sim 20$ so that $T_{\hbox{f}} >$ GeV, 
we can use $g_{*{\hbox{f}}}\simeq 106.75$ \cite{Torrente} in equations 
(\ref{eqxf}) -- (\ref{eqYinf}). Equation (\ref{eqxf}) shows how $\xf$ has 
a logarithmic dependence on $m$, while theoretical considerations 
\cite{Ellis,Report,Torrente,Roszkowski,Fornengo,Ellis2} related to the 
minimal supersymetric extensions of the Standard Model (MSSM) yield specific 
forms for $a$ and $b$ that also depend on $m$. Inserting into 
(\ref{eqOmega0})--(\ref{eqYinf}) the specific forms of $a$ and $b$ for each 
annihilation channel leads to a specific range of $m$ that satisfies the 
``abundance'' criterion based on current observational constraints that 
require $0.1 < \Omega_0 < 0.3$ and $h\approx 0.65$ \cite{Peac}.

Suitable forms for $\la \sigma |v|\ra$ can be obtained for all types of
annihilation reactions \cite{Report}. If the neutralino is mainly pure B-ino, 
it will mostly annihilate into lepton pairs through t-channel exchange of 
right-handed sleptons. In this case the cross section is p-wave dominated 
and can be approximated by (\ref{eq<sv>}) with \cite{Torrente,Moroi,Olive}
\ba 
a \ \approx \ 0,\qquad b \ \approx
\ \frac{8\,\pi\,\alpha_1^2}{m^2\,\left[1+m_l^2/m^2\right]^2},
\label{sleptons}
\ea
where $m_{l}$ is the mass of the right-handed slepton ($m_{l} \sim m$ 
\cite{Torrente}) and $\alpha_1^2 = g_1^2/4 \pi \simeq 0.01$ is the fine 
structure coupling constant for the $U(1)_Y$ gauge interaction. If the 
neutralino is Higgsino-like, annihilating  into W-boson pairs, then the cross 
section is s-wave dominated and can be approximated by (\ref{eq<sv>}) with 
\cite{Torrente,Moroi,Olive}
\ba 
b \ \approx \ 0,\qquad a \ \approx \
\frac{\pi\,\alpha_2^2\,(1-m_{ W}^2/m^2)^{3/2}}
{2\,m^2\,(2-m_{_W}^2/m^2)^2},
\label{Wboson}
\ea
where $m_{_W}=80.44$ GeV is the mass of the W-boson and 
$\alpha_2^2 = g_2^2/4 \pi \simeq 0.03$ is the fine structure coupling constant 
for the $SU(2)_L$ gauge interaction.      

In the freeze out era the entropy per particle (in units of the Boltzmann 
constant $\kB$) for the neutralino gas is given by \cite{KoTu,Peac,Padma1}
\ba 
\sff \ = \ \left[\frac{\rho + p}{n\,T}\right]_{\hbox{f}} \ = \
\frac{5}{2} \ + \ \xf,
\label{sf}
\ea
where we have assumed that chemical potential is negligible and have used
the equation of state (\ref{eqst}). From (\ref{eqxf}) and (\ref{sf}), it is 
evident that the dependence of $\sff$ on $m$ will be determined by the 
specific details of the annihilation processes through the forms of $a$ and 
$b$. In particular, we will use (\ref{sleptons}) and (\ref{Wboson}) to compute
$\sff$ from (\ref{eqxf})-(\ref{sf}).

\section{The microcanonical entropy}

After the freeze out era, particle numbers are conserved and the neutralinos 
constitute a weakly interacting and practically collision--less 
self--gravitating gas. This gas is only gravitationally coupled to other 
components of the cosmic fluid. As it expands, it experiences free streaming 
and eventually undergoes gravitational clustering forming stable bound 
virialized structures \cite{Peac,Padma1,Padma2,Padma3}. The evolution between 
a spectrum of density perturbations at the freeze out and the final virialized 
structures is extremely complex, involving a variety of dissipative effects 
characterized by collisional and collision--less relaxation processes 
\cite{Padma2,Padma3,HD}.  However, the freeze out and present day virialized 
structures roughly correspond to ``initial'' and ``final'' equilibrium states 
of this gas. Therefore, instead of dealing with the enormous complexity of 
the details of the intermediary processes, we will deal only with quantities 
defined in these states with the help of simplifying but general physical 
assumptions. 

The microcanonical ensemble in the ``mean field''  approximation yields an 
entropy definition that is well defined for a self--gravitating gas in an 
intermediate scale, between the short range and long range regimes of the 
gravitational potential. This intermediate scale can be associated with a 
region that is ``sufficiently large as to contain a large number of particles 
but small enough for the gravitational potential to be treated as a 
constant''~\cite{Padma2}. Considering the neutralino gas in present day
virialized halo structures as a diluted, non-relativistic (nearly) ideal gas 
of weakly interacting particles, its microcanonical entropy per particle 
under these conditions can be given in terms of the volume of phase space
\cite{Padma3} 
\ba 
s \ = \ \ln \,\left[\frac{\,(2mE)^{3/2}\,V\,}{(2\pi\hbar)^3}\right],
\label{mcsdef}
\ea
where $V$ and $E$ are local average values of volume and energy associated 
with the intermediate scale. For  non-relativistic velocities $v/c\ll 1$, we 
have $V\propto 1/n\propto m/\rho$ and $E\propto m\,v^2/2\propto m/x$. In fact, 
under these assumptions the definition (\ref{mcsdef}), evaluated at the freeze 
out, is consistent with (\ref{n_theq}) and (\ref{sf}), and so it is also valid 
immediately after the freeze out era (once particle numbers are conserved). 
Since (\ref{mcsdef}) is valid at both the initial and final states, 
respectively corresponding to the decoupling ($\sff,\,\xf,\,\nf$) and the 
values ($\sha,\,\xha,\,\nha$) associated with a suitable halo structure, the 
change in entropy per particle that follows from (\ref{mcsdef}) between these 
two states is given by
\ba 
\sha \ - \ \sff \ = \ 
\ln\,\left[\frac{\nf}{\nha}\left(\frac{\xf}{\xha}\right)^{3/2}
\right],
\label{Delta_s}
\ea
where (\ref{sf}) can be used to eliminate $\sff$ in terms of $\xf$. 
Considering present day halo structures as roughly spherical, inhomogeneous 
and self-gravitating gaseous systems, the intermediate scale of the 
microcanonical description is an excellent approximation for gas particles 
in a typical region of $\sim 1\,\hbox{pc}^3$ within the halo core, near the 
symmetry center of the halo where the gas density enhancement is maximum but 
spacial gradients of all macroscopic quantities are negligible~\cite{HG,IS2}. 
Therefore, we will consider current halo macroscopic variables as evaluated 
at the center of the halo: $\shac,\,\xhac,\,\nhac$.      

In order to obtain a convenient theoretical estimate of $\shac$ from 
(\ref{Delta_s}), we need to relate $\nf$ with present day cosmological 
parameters like $\Omega_0$ and $h$. Bearing in mind that density 
perturbations at the freeze out era were very small 
($\delta\,\nf/\nf < 10^{-4}$, \cite{KoTu,Padma1,Peac}), the density $\nf$ is 
practically homogeneous and so we can estimate it from the conservation of 
particle numbers: $\nf = n_0 \,(1+z_{\hbox{f}})^3$, and of photon entropy: 
$g_{*\hbox{f}}\CS_{\hbox{f}} = g_{*0}\,\CS_0\, \,(1+z_{\hbox{f}})^3$, valid 
from the freeze out era to the present for the unperturbed homogeneous 
background. Eliminating $ (1+z_{\hbox{f}})^3$ from these conservation laws 
yields
%
\ba 
\nf \ = \
n_0\,\frac{g_{*\hbox{f}}}{g_{*0}}\left[\frac{T_{\hbox{f}}}
{T_0^{\hbox{\tiny{CMB}}}}\right]^3
\ \simeq \ 27.3\,n_0\,
\left[\frac{x_0^{\hbox{\tiny{CMB}}}}{\xf}\right]^3,\label{eqnf}\\
\hbox{where}\quad x_0^{\hbox{\tiny{CMB}}} \ \equiv \
\frac{m}{T_0^{\hbox{\tiny{CMB}}}}
\ = \ 4.29\,\times\,10^{12}\,\frac{m}{\hbox{GeV}}\nonumber
\ea
%
where $g_{*0}=g_*(T_0^{\hbox{\tiny{CMB}}})\simeq 3.91$ and
$T_0^{\hbox{\tiny{CMB}}}=2.7\,\hbox{K}$. Since for present day conditions 
$n_0/\nhac=\rho_0/\rhohac$ and $\rho_0=\rho_{\hbox{crit}}\,\Omega_0\,h^2 $, 
we collect the results from (\ref{eqnf}) and write (\ref{Delta_s}) as 
\ba 
\shacth = \xf + 93.06 +
\ln\left[\left(\frac{m}{\hbox{GeV}}\right)^3\,\frac{h^2\,\Omega_0}
{(\xf\,\xhac)^{3/2}}\,\frac{\rho_{\hbox{crit}}}{\rhohac}\right]\nonumber\\
= \xf + 81.60 +
\ln\left[\left(\frac{m}{\hbox{GeV}}\right)^3\,\frac{h^2\,\Omega_0}
{(\xf\,\xhac)^{3/2}}\,\frac{\hbox{GeV/cm}^3}{\rhohac}\right],
\label{shalo}
\ea 
Therefore, given $m$ and a specific form of $\la\sigma|\hbox{v}|\ra$ 
associated with $a$ and $b$, equation (\ref{shalo}) provides a theoretical 
estimate of the entropy per particle of the neutralino halo gas that depends 
on the initial state given by $\xf$ in (\ref{eqxf}) and (\ref{sf}), on 
observable cosmological parameters $\Omega_0,\,h$ and on generic state 
variables associated to the halo structure.

\section{Theoretical and empiric entropies} 

If the neutralino gas in present halo structures strictly satisfies MB 
statistics, the entropy per particle, $\shac$, in terms of $\rhohac=m\,\nhac$ 
and $\xhac=m\,c^2/(\kB\,\Tha_c)$, follows from  the well known Sackur--Tetrode 
entropy formula \cite{Pathria}
\ba 
&&\shacmb = 
\frac{5}{2}+\ln\left[\frac{m^4\,c^3}{\hbar^3\,(2\pi\,\xhac)^{3/2}\,
\rhohac}\right]
\nonumber\\ &&= 94.42 +
\ln\left[\left(\frac{m}{\hbox{GeV}}\right)^4\,
\left(\frac{1}{\xha_c}\right)^{3/2}
\,\frac{\hbox{GeV/cm}^3}{\rhohac}\right].
\label{s_halo}
\ea
Such a MB gas in equilibrium is equivalent to an isothermal halo if we
identify \cite{BT} 
\ba 
\frac{c^2}{\xha} \ = \ \frac{\kB\,\Tha}{m} \ 
= \ \sigha^2,
\label{isot_MBa}
\ea
where $\sigha^2$ is the velocity dispersion (a constant for isothermal halos). 

However, an exactly isothermal halo is not a realistic model, since its total 
mass diverges and it allows for infinite particle velocities (theoretically 
accessible in the velocity range of the MB distribution). More realistic halo 
models follow from ``energy truncated'' (ET) distribution functions 
\cite{Padma3,BT,Katz1,Katz2,MPV} that assume a maximal ``cut off'' velocity 
(an escape velocity). Therefore, we can provide a convenient empirical 
estimate of the halo entropy, $\shac$, from the microcanonical entropy 
definition (\ref{mcsdef}) in terms of phase space volume, but restricting this 
volume to the actual range of velocities (\hbox{i.e.} momenta) accessible to 
the central particles, that is up to a maximal escape velocity $v_e(0)$. From 
theoretical studies of dynamical and thermodynamical stability associated 
with ET distribution functions \cite{Katz1,Katz2,cohn,MPV,HM,GZ, RST} and from 
observational data for elliptic and LSB galaxies and clusters
\cite{young,DBM,HG,FDCHA1,FDCHA2}, it is reasonable to assume
\ba 
v_e^2(0) \ = \ 2\,|\Phi(0)| \ \simeq \ \alpha \, \sigha^2(0),\quad 12<
\alpha< 18,
\label{alphas}
\ea
where $\Phi(r)$ is the newtonian gravitational potential. We have then   
\ba 
&&\shacem \ \simeq \
\ln\left[\frac{m^4\,v_{e}^3}{(2\pi\hbar)^3\,\rhohac}\right]\nonumber\\ 
=&& 89.17 +\ln\left[\left(\frac{m}{\hbox{GeV}}\right)^4\,
\left(\frac{\alpha}{\xha_c}\right)^{3/2}
\,\frac{\hbox{GeV/cm}^3} {\rhohac}\right],
\label{SHALO}
\ea    
where we used $\xhac=c^2/\sigha^2(0)$ as in (\ref{isot_MBa}). As expected, 
the scalings of (\ref{SHALO}) are identical to those of (\ref{s_halo}). 
Similar entropy expressions for elliptic galaxies have been examined in 
\cite{LGM}.

Comparison between $\shac$ obtained from (\ref{SHALO}) and from (\ref{shalo}) 
leads to the constraint 
\ba 
\shacth &&= \ \shacem \quad \Rightarrow\nonumber\\ \xf && = \ 7.57 \ + \
\ln\left[\frac{(\alpha\,\xf)^{3/2}}{h^2\,\Omega_0}\,
\frac{m}{\hbox{GeV}}\right].
\label{constr}
\ea
which does not depend on the halo variables $\xhac,\,\rhohac$, hence it can 
be interpreted as the constraint on $\sff=5/2+\xf$ that follows from the 
condition $\shacth = \ \shacem$. Since we can use (\ref{eqOmega0}) and 
(\ref{eqYinf}) to eliminate $h^2\,\Omega_0$, the constraint (\ref{constr}) 
becomes a relation involving only $\xf,\,m,\,a,\,b,\,\alpha$. This constraint 
is independent of  (\ref{eqxf}), which is another (independent) expression for 
$\sff=5/2+\xf$, but an expression that follows {\it{only}} from the neutralino 
annihilation processes. Therefore, the comparison between $\shacth$ and 
$\shacem$, leading to a comparison of two independent expressions for $\sff$, 
is not trivial but leads to  an ``entropy consistency'' criterion that can be 
tested on suitable desired values of $m,\,a,\,b,\,\alpha$. This implies that 
a given dark matter particle candidate, characterized by $m$ and by specific 
annihilation channels given by $\xf$ through (\ref{eqxf}), will pass or fail 
to pass this consistency test independently of the details one assumes 
regarding the present day dark halo structure. This is so, whether we conduct 
the consistency test by comparing (\ref{eqxf}) and (\ref{constr}) or 
(\ref{shalo}) and (\ref{SHALO}). However, the actual values of $\shac$ for a 
given halo structure, whether obtained from (\ref{SHALO}) or from 
(\ref{shalo}), do depend on the precise values of $\rhohac$ and $\xha_c$.
Since the matching of either (\ref{eqxf}) and (\ref{constr}) or (\ref{shalo}) 
and (\ref{SHALO}) shows a weak logarithmic dependence on $m$, the fulfillment 
of the  ``entropy consistency'' criterion identifies a specific mass range 
for each dark matter particle. This allows us to  discriminate, in favor or 
against, suggested dark matter particle candidates and/or annihilation 
channels by verifying if the standard abundance criterion (\ref{eqOmega0}) 
is simultaneously satisfied for this range of masses. 

\section{Testing the entropy consistent criterion}   

Since we can write (\ref{constr}) as:
\ba
\ln(h^{2} \Omega_0) = 7.57 - \xf + \ln\left[(\alpha \xf)^{3/2} m \right].
\label{Result}
\ea
this constraint becomes a new estimate of the cosmological parameters 
$h^{2} \Omega_0$, given as in terms of a structural parameter of galactic 
dark matter halos, $\alpha$, the mass of the neutralino, $m$, and the 
temperature of the neutralino gas at freeze out, $\xf$. This last quantity 
depends explicitly not only on $m$, but also on its interaction cross section, 
and hence on the details of its phenomenological physics {\it viz} 
(\ref{eqxf}).

At this point we consider values for the constants $a$ and $b$ that define the
interaction cross section of the neutralino, and use (\ref{Result}) to plot 
$\Omega_0$ as a function of $m$ in GeV's. Using $h=0.65$ and given the
uncertainty range of $\alpha$, we will obtain not a curve, but a region in 
the $\Omega_0-m$ plane. Considering first condition (\ref{Wboson}), 
corresponding to Higgsino--like neutralinos, leads to the shaded region in 
figure 1a. On this figure we have also plotted the relation which the 
abundance criterion (\ref{eqOmega0}) yields on this same plane. Firstly, we 
notice that the mass range that results from our entropy criterion intersects 
the one resulting from the  abundance criterion. However, it is evident that
within the observationally determined range of $\Omega_0$ (the horizontal 
dashed lines 0.2-0.4), there is no intersection between the shaded region and 
the abundance criterion curve.  This implies that both criteria are mutually 
inconsistent, thus the possibility that Higgsino-like neutralinos make up both 
the cosmological dark matter and galactic dark matter appears unlikely.

Repeating the same procedure for mainly B--ino neutralinos, (\ref{sleptons}) 
yields figure 1b. In this case, we can see that the abundance criterion curve 
falls well within the shaded region defined by the entropy criterion. Although 
we can not improve on the mass estimate provided by the abundance criterion 
alone, the consistency of both criteria reveals the B-ino neutralino as a 
viable option for both the cosmological and the galactic dark matter.

\begin{figure}
\centering
\includegraphics[height=5.7cm]{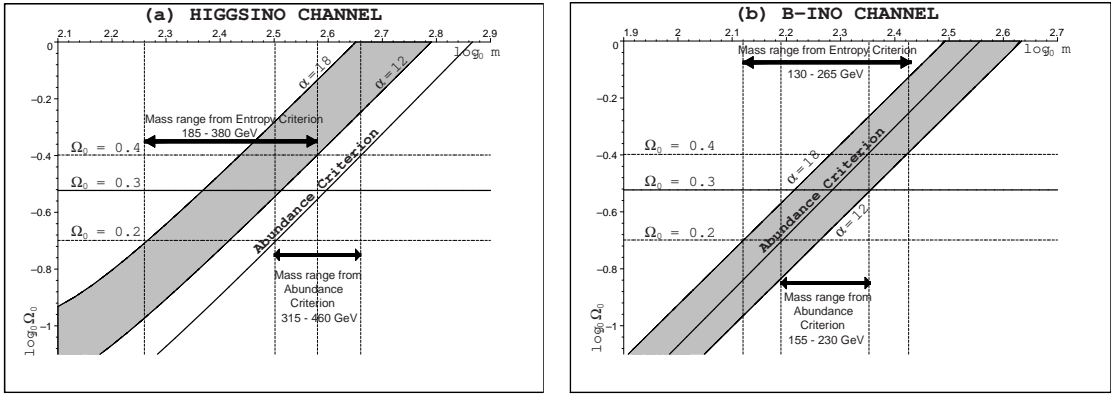}
%
%
\caption{Figures (a) and (b) respectively correspond to the Higgsino and 
B-ino channels. The shaded regions display $\Omega_0$ vs m from our entropy 
criterion (\ref{Result}) with the solid curve giving $\Omega_0$ from the 
cosmological abundance criterion (\ref{eqOmega0}), in all cases for $h=0.65$.
The horizontal dashed lines give current estimates of $\Omega_0=0.3\pm0.1$. 
It is evident that only the B-ino channels allow for a simultaneous fitting of 
both the abundance and the entropy criteria.}
\label{fig:2}       
\end{figure}

It is also interesting to evaluate (\ref{SHALO}) and (\ref{shalo}) for the 
two cases of neutralino channels: the B-ino and Higgsino, but now considering 
numerical estimates for $\xha$ and $\rhoha$ that correspond to central regions 
of actual halo structures. Considering terminal velocties in rotation curves 
we have 
$v_{\hbox{term}}^2\simeq 2 \sigha^2(0)$, so that 
$\xhac \simeq 2(c/v_{\hbox{term}})^2$, while recent data from LSB
galaxies and clusters \cite{FDCHA1, FDCHA2, IS1, IS2, IS3} suggest the range 
of values
$0.01\,\hbox{M}_\odot/\hbox{pc}^3 < \rhohac< 1\,\hbox{M}_\odot/\hbox{pc}^3$. 
Hence, we will use in the comparison of (\ref{shalo}) and (\ref{SHALO}) the 
following numerical values: $\rhohac=0.01\,\hbox{M}_\odot/\hbox{pc}^3 =0.416
\,\hbox{GeV/cm}^3$ and $\xhac = 2\times 10^6$, typical values for a large 
elliptical or spiral galaxy with  $v_{\hbox{term}}\simeq 300\,\hbox{km/sec}$ 
\cite{IS1,IS2,IS3}.
\begin{figure}
\centering
\includegraphics[height=9.5cm]{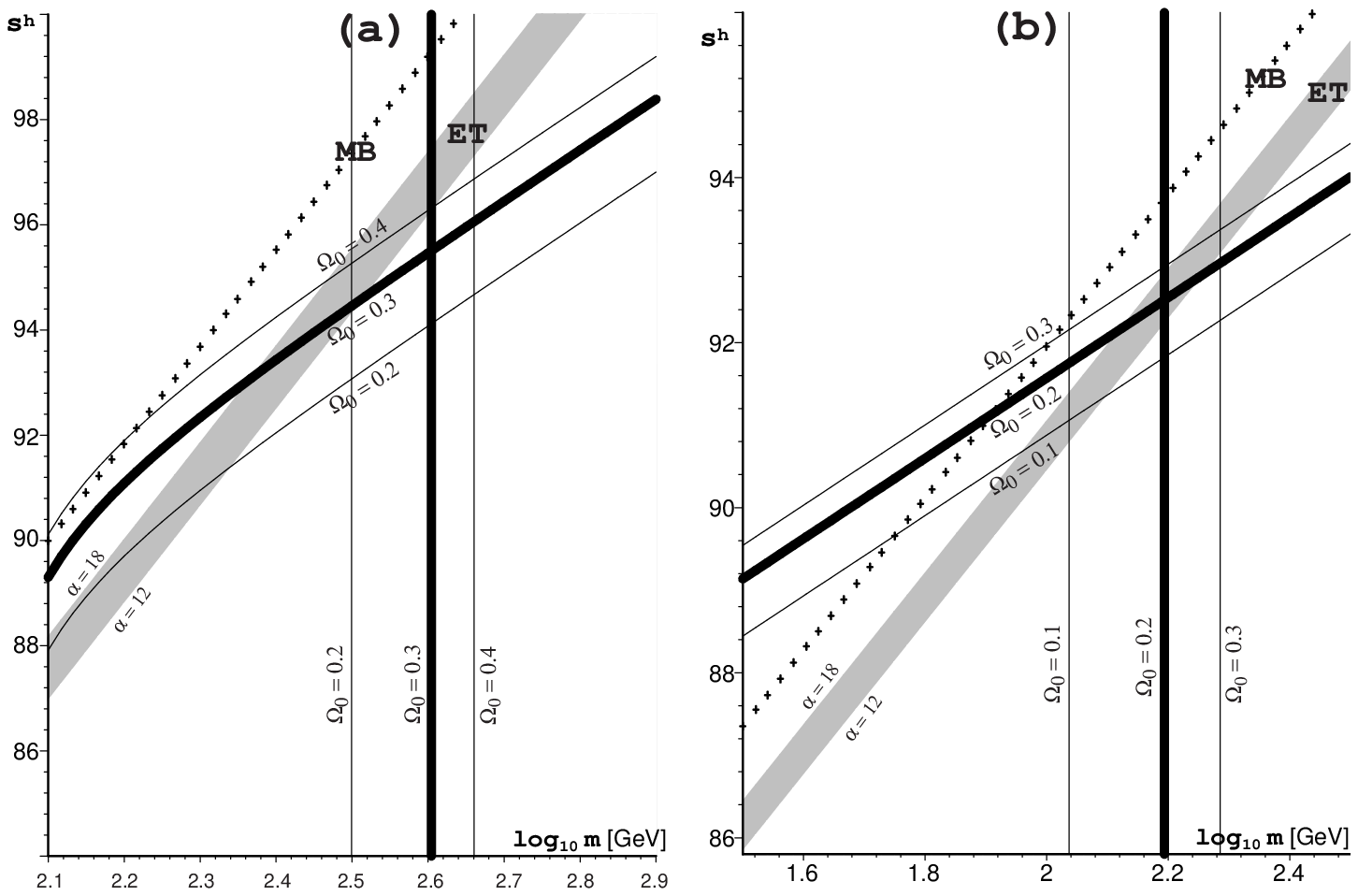}
%
\caption{Figures (a) and (b) respectively correspond to the Higgsino and 
B-ino channels. The figures display $\shacem$ from 
(\ref{alphas})--(\ref{SHALO}) (gray strip), $\shacth$ from
(\ref{shalo}) for $h=0.65$ and the uncertainty strip $\Omega_0=0.3\pm0.1$ 
(thick curves) and $\shacmb$ from (\ref{s_halo}) (crosses), all of them as 
functions of $\log_{10} m$. The vertical strip marks the range of values 
of $m$ that follow from (\ref{eqOmega0})--(\ref{eqYinf}) for the same values 
of $\Omega_0$ and $h$. It is evident that only the B-ino channels allow 
for a simultaneous fitting of both the abundance and the entropy criteria.}
\label{fig:1}       
\end{figure}
Figure 2a displays $\shacth$ and $\shacem$ as functions of $\log_{10}\,m$, 
for the halo structure described above, for the case of a neutralino that is 
mostly Higgsino. The shaded region marks $\shacem$ given by (\ref{SHALO}) 
for the range of values of $\alpha$, while the vertical lines correspond to 
the range of masses selected by the abundance criterion (\ref{eqOmega0}) for 
$\Omega_0=0.2,\,0.3,\,0.4$. The solid curves are $\shacth$ given by 
(\ref{shalo}) for the same values of $\Omega_0$, intersecting the shaded 
region associated with (\ref{SHALO}) at some range of masses. However, the 
ranges of coincidence of a fixed (\ref{shalo}) curve with the shaded region
(\ref{SHALO}) occurs at masses which correspond to values of $\Omega_0$ that 
are different from those used in (\ref{shalo}), that is, the vertical lines 
and solid curves with same $\Omega_0$ intersect out of the shaded region. 
Hence, this annihilation channel does not seem to be favored.

Figure 2b depicts the same variables as figure 2a, for the same halo 
structure, but for the case of a neutralino that is mostly B-ino. In this 
case, the joint application of the abundance and entropy criteria yield a 
consistent mass range of 
$150\,\hbox{GeV} < m_{{\tilde\chi^1_0}} < 250 \,\hbox{GeV}$), which  allows 
us to favor this annihilation channel as a plausible dark matter candidate, 
with $m$ lying in the narrow ranges given by this figure for any chosen value 
of $\Omega_0$. As noted above, the results of figures 1a and 1b are totally 
insensitive to the values of halo variables, $\xhac$ and $\rhohac$, used in 
evaluating (\ref{SHALO}) and (\ref{shalo}). Different values of these 
variables (say, for a different halo structure) would only result in a 
relabeling of the values of $\shac $ along the vertical axis of the figures.

\section{Conclusions} 

We have presented a robust consistency criterion that can be verified for any
annihilation channel of a given dark matter candidate proposed as the
constituent particle of the present galactic dark matter halos. Since we 
require that the empirical estimate $\shacem$ of present dark matter haloes 
must match the theoretical value $\shacth$, derived from the microcanonical 
definition and from freeze out conditions for the candidate particle, the 
criterion is of a very general applicability, as it is largely insensitive to
the details of the structure formation scenario assumed. Further, the details 
of the present day halo structure enter only through an integral feature of 
the dark halos, the central escape velocity, thus our results are also 
insensitive to the fine details concerning the central density and the various 
models describing the structure of dark matter halos. A crucial feature of this
criterion is its direct dependence on the physical details ({\it{i.e.}}
annihilation channels and mass) of any particle candidate.  

Recent theoretical work by  E. A. Baltz {\it{et al.}} \cite{HEAT-TH} confirmed
that neutralino annihilation in the galactic halo can produce enough positrons 
to make up for the excess of cosmic ray positrons experimentally detected by 
the HEAT collaboration \cite{HEAT1,HEAT2}. Baltz {\it{et al.}} concluded that 
for a boost factor $B_s \sim 30$ the neutralinos must be primarily B-inos with 
mass around 160 GeV. For a boost factor $30 < B_s < 100$, the 
gaugino--dominated SUSY models complying with all constraints yield neutralino 
masses in the range of 
$150\,\hbox{GeV} <  m_{{\tilde\chi^1_0}} < 400 \,\hbox{GeV}$. On the other 
hand, Higgsino dominated neutralinos are possible but only for $B_s \sim 1000$ 
with masses larger than 2 TeV. The results that we have presented in this 
paper are in agreement with the predictions that follow from \cite{HEAT-TH}, 
as we obtain roughly the same mass range for the B-ino dominated case (see 
figure 1b) and the Higgsino channel is shown to be less favored in the mass 
range lower than TeV's.       

We have examined the specific case of the lightest neutralino for the mostly
B-ino and mostly Higgsino channels. The joint application of the 
``entropy consistency'' and the usual abundance criteria clearly shows that 
the B-ino channel is favored over the Higgsino. This result can be helpful 
in enhancing the study of the parameter space of annihilation channels of 
LSP's in  MSSM models, as the latter only use equations (\ref{eqxf}) and 
(\ref{eqOmega0})--(\ref{eqYinf}) in order to find out which parameters yield 
relic gas abundances that are compatible with observational constraints 
\cite{Ellis,Report,Torrente,Roszkowski,Fornengo,Ellis2}. However, equations 
(\ref{eqxf}) and (\ref{eqOmega0})--(\ref{eqYinf}) by themselves are 
insufficient to discriminate between annihilation channels. A more efficient 
study of the parameter space of MSSM can be achieved by the joint usage of 
the two criteria, for example, by considering more general cross section 
terms (see for example \cite{Report}) than the simplified approximated forms 
(\ref{sleptons}) and (\ref{Wboson}). This work is currently in progress.\\

\begin{theacknowledgments}
This work has been supported in part by grants; {\bf PAPIIT} project No. 
{\tt IN109001} and in part by {\bf CoNaCyT} projects No. {\tt I37307-E} and 
No. {\tt I39181-E}.
\end{theacknowledgments}

\end{document}